\documentclass[preprint2,epsfig]{aastex}

\usepackage{amsmath}
\usepackage{hyperref}
\usepackage{graphics}
\usepackage{natbib}
\title{The Changing Fractions of Type Ia Supernova NUV-Optical \\
Subclasses with Redshift}

\author{Peter~A.~Milne\altaffilmark{1},
Ryan~J.~Foley\altaffilmark{2,3}, \\
Peter~J.~Brown\altaffilmark{4},
Gautham Narayan\altaffilmark{5} \\
\altaffiltext{1}{University of Arizona, Steward Observatory,
933 N. Cherry Ave., Tucson, AZ 85719, USA; pmilne@as.arizona.edu}
\altaffiltext{2}{
Astronomy Department,
University of Illinois at Urbana-Champaign,
1002 West Green Street,
Urbana, IL 61801, USA; rfoley@illinois.edu
}
\altaffiltext{3}{
Department of Physics,
University of Illinois Urbana-Champaign,
1110 West Green Street,
Urbana, IL 61801, USA
}
\altaffiltext{4}{George P. and Cynthia Woods Mitchell Institute for Fundamental Physics \& Astronomy,
Texas A. \& M. University, Department of Physics and Astronomy,
4242 TAMU, College Station, TX 77843, USA;
pbrown@physics.tamu.edu}
\altaffiltext{5}{National Optical Astronomy Observatory, P.O. Box 26732, Tucson, AZ 85726, 
USA; gnarayan@noao.edu}
}

\newcommand{\kms}{$km s^{-1}$}

\begin{document}
\begin{abstract}

  UV and optical photometry of Type Ia supernovae (SNe~Ia) at low
  redshift have revealed the existence of two distinct color groups, 
  NUV-red and NUV-blue events. The color curves differ primarily by an offset, 
  with the NUV-blue $u-v$ color curves bluer than the 
  NUV-red curves by 0.4 mag. For a sample of 23 low-$z$  
  SNe~Ia observed with {\it Swift}, the NUV-red group dominates by a ratio of 2:1. 
  We compare rest-frame UV/optical spectrophotometry of intermediate and 
  high-$z$ SNe~Ia with UVOT photometry and {\it HST} spectrophotometry of
  low-$z$ SNe~Ia, finding that the same two color groups exist at
  higher-$z$, but with the NUV-blue events as the
  dominant group. Within each red/blue group, we do not detect any offset in color for 
  different redshifts, providing insight into how SN~Ia UV emission evolves with redshift. 
  Through spectral comparisons of SNe~Ia with similar peak widths 
  and phase, we explore the wavelength range that produces the UV/OPT color 
  differences. 
  We show that the ejecta velocity of NUV-red SNe is larger
  than that of NUV-blue objects by roughly 12\% on average.  This
  velocity difference can explain some of the UV/optical color difference, but
  differences in the strengths of spectral features seen in mean
  spectra require additional explanation.  
  Because of the different $b-v$ colors for these groups, NUV-red SNe 
  will have their extinction underestimated using common techniques.
  This, in turn, leads to under-estimation of the optical
  luminosity of the NUV-blue SNe~Ia, in particular, for the
  high-redshift cosmological sample.  Not accounting for this effect
  should thus produce a distance bias that increases with redshift and
  could significantly bias measurements of cosmological parameters.

\end{abstract}

\vspace{4mm}
\section{Introduction}

Type Ia supernovae (SNe~Ia) are very luminous with a well-understood
variation of their peak optical luminosities as a function of their
light-curve shapes, the width-luminosity relation (WLR;
$\Delta$m$_{15}$(B): Phillips et al. 1993; $\Delta$: Riess et al.
(1996), Jha et al. (2006a); stretch(s): Perlmutter et al. (1997); 
SiFTO: Conley et al. (2008)). The
empirically derived width-luminosity relation has allowed SNe~Ia to be
utilized to study the expansion of the universe, revealing that the
expansion is currently experiencing a period of acceleration 
(Riess et al. 1998; Perlmutter et al. 1999). In order to
lower the scatter in distance determinations, the larger sample of all
SNe~Ia is reduced to a group of so-called ``normal'' SNe~Ia, a group
that follow a regular variation of their properties with the width of
the peak of their optical light curves (i.e., scaling).  The reduced
sample eliminates narrow-peaked SNe~Ia (see Taubenberger et al. 2008;
Krisciunas et al.  2009), SNe~Iax (Li et al. 2003; Foley et al.
2013), candidates for super-Chandrasekhar SNe~Ia (e.g., Howell et al.
2006; Scalzo et al.  2010) and individually anomalous events
\citep[e.g.,][]{Li_etal_2001,Foley10:06bt,Maguire_etal_2010, Sullivan_etal_2011}.  
The reduced sample comprises roughly
half of the SNe~Ia discovered, and will be the focus of this work.

A key element of the cosmological utilization of SNe~Ia is the assumption
that the SN~Ia phenomenon does not change as the universe has aged, or
that any changes continue to follow the empirical laws derived for
nearby normal SNe~Ia.  For example, the method can work if higher-$z$
SNe~Ia tend to have broader optical light curves, as long as the 
luminosities of those
SNe still follow the WLR for that peak width.  The problem would arise
if higher-$z$ SNe~Ia had brighter or fainter luminosities for the same
light-curve shape. An example of one such problem was reported by 
Sullivan et al. (2010), who claim a dependence of SN~Ia luminosity on the 
mass and specfic star formation rates of the host galaxy. As these 
characteristics vary with redshift, a systematic error is introduced to 
cosmological analyses at the 0.08 magnitude level. 

Spectral modeling makes it clear that iron-peak elements significantly
impact the ultraviolet (UV) spectral-energy distribution (SED) of
SNe~Ia \citep[e.g.,][]{Kirshner_etal_1993, Hoflich_etal_1998, Lentz_etal_2000, 
Sauer_etal_2008, Mazzali_etal_2013, Hachinger_etal_2013}.
Theoretical models of the thermonuclear explosion of a SN~Ia have 
explored the effects of varying the progenitor metallicity or explosion 
physics on the
resulting $^{56}$Ni production and distribution, as well as on the
expansion velocities (H\H{o}flich et al. 1998; Lentz et al. 2000;
Timmes et al. 2003; Sauer et al. 2008; Walker et al. 2012). All
simulations suggest that the UV emission is much more strongly
affected by metallicity variations than the optical emission, and
would lead to variations in the UV-optical colors of SNe~Ia with
different progenitor metallicities.  Since the progenitor metallicity
is one aspect of a SN~Ia explosion that would be expected to change
with redshift, it is vital to establish the nature of the progenitor
metallicity -- UV SED correlation.  However, there is not a consensus
about the magnitude of that correlation, underscoring how
complicated the challenge is to model the complete time-evolving
spectrum of a SN~Ia.  Further, as there are multiple classes of
progenitor systems suggested (e.g., single versus double degenerate),
as well as multiple explosion scenarios (e.g., delayed detonations
versus deflagrations versus double detonations), it is not clear that
progenitor metallicity differences will be the dominant factor in
variations of the UV-optical SED \citep{Roepke12,foley12b}.
With all the challenges facing 
theoretical modeling, observational efforts to characterize the time
evolution of the UV-optical SED for a large sample of SNe~Ia might
guide the theoretical investigations.

The importance of UV emission as a probe of the SN~Ia event has been
recognized for decades, but it has been challenging to perform
rest-frame UV observations.  Efforts to observe higher-$z$ SNe~Ia has
increased in recent years with many large telescopes being utilized.
Ellis et al. (2008) used the LRIS spectrograph on the Keck I telescope
to observe Supernova Legacy Survey (SNLS) SNe~Ia, using SNLS
photometry to remove host galaxy light from the spectra (hereafter
E08).  E08 found that the dispersion of individual spectra about a
mean mid-$z$ spectrum increased in the UV. The mean spectrum was not
found to have evolved compared to a low-$z$ mean spectrum.  Foley et
al. (2008) performed an independent study that used a collection of
large telescopes to observe ESSENCE mid-$z$ SNe~Ia, also finding that
the spectral dispersion increased to the UV.  However, uncertain
galaxy contamination correction prevented a robust comparison of
spectral continua.  
%\citet{Foley12:sdss} 
Foley et al. (2012) employed the LRIS spectrograph on Keck to
observe 21 mid-$z$ SNe~Ia from the Sloan Digital Sky Survey
Supernova Survey (hereafter F12).  F12 was able to remove galaxy
contamination from the spectra by modeling the galaxy SED from SDSS
photometry, and used a low-$z$ sample of 
optical SN~Ia spectra 
\citep*{Silverman_Kong_Filippenko_2012} 
combined with a sample of 46 low-$z$ UV spectra \citep{foley08}.
With this high-quality data set, F12 found that while
the optical spectra were nearly identical, the UV portion of a mean
spectrum of the mid-$z$ Keck/SDSS sample had excess flux relative to a
low-$z$ mean spectrum.

Sullivan et al. (2009) compared a low-$z$ mean spectrum, the E08
 mid-$z$ mean spectrum, and a high-$z$ [$z \geq 0.6$ SNe
with {\it HST}/ACS spectra; 
% XXXX
%\citet{Riess_etal_2007}] 
Riess et al. (2007)] mean spectrum.  There
was an indication for UV flux excess for the higher redshift samples,
but this was discounted because of the relatively small low-$z$ sample
used in the analysis.  Cooke et al. (2011) and Maguire et al. (2012) expanded
the low-$z$ sample, paying careful consideration to selection effects;
again, there were differences in the UV continuum. 

A study by Balland et al. (2009) used the ESO/VLT telescope to observe
SNLS SNe~Ia, using CFHT photometry to remove host galaxy light from
the spectra (hereafter B09).  That work concentrated on the variation
of spectral features with redshift, finding that higher-$z$ spectra
have shallower intermediate mass element (IME) absorption features,
but this was attributed to different color distributions for
the samples, as the stretch values were similar given the uncertainties.  
Collectively, these studies show that there is little
evolution of the optical emission with redshift, but more in the UV.
The direction of that evolution is for the higher-$z$ SNe~Ia to be
brighter in the UV.  At a given redshift, the UV emission exhibits
increased dispersion compared to the optical wavelength range. One
interpretation of these findings has been improved confidence for the
cosmological utility of SNe~Ia in the optical wavelength range,
combined with skepticism for the cosmological utility of the UV
wavelength range.

In Section 2, we compare the $u-v$ colors of low-$z$ versus mid- and high-$z$ 
photometry and spectrophotometry. In Section 3, we compare paired spectra from 
each group. In Section 4, we compare $u-b$ and $b-v$ colors, with Section 5 
devoted to exploring the cosmological implications of the optical color 
differences. Section 6 summarizes these findings and argues for altering 
light curve fitting routines to account for the existence of two color groups. 

\section{Comparing $u-v$ colors of low-, mid- \& high-$z$ Samples}

\begin{figure}[h]
\epsscale{1.0} \plotone{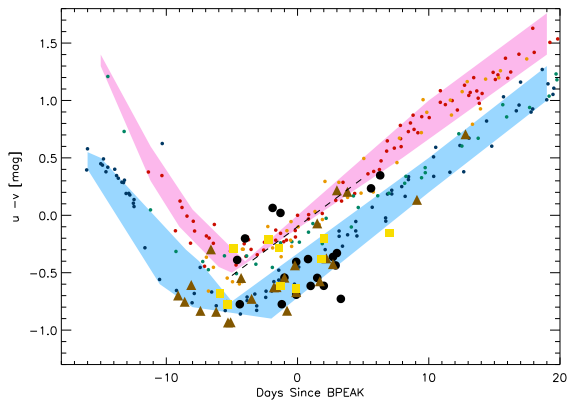}
\caption{$u-v$ colors of low-$z$ versus mid-$z$ samples. UVOT photometry is shown with small circles, 
and is contained in blue and red shaded regions representing the range of colors within each group. 
Foley et al. (2012) mid-$z$ spectrophotometry is shown with large, black circles. Ellis et al. (2008) 
mid-$z$ spectrophotometry is shown with large, brown triangles. Balland et al. (2009) mid-$z$ 
spectrophotometry is shown with large, yellow squares. UVOT photometry is color-coded according to 
M13, with NUV-blue (blue), NUV-red (red), MUV-blue (orange) and irregular (green). The mid-$z$ 
photometry exhibits separation into two colors curves, as seen in the UVOT photometry in M13.}
\label{u_v_color_curve}
\end{figure}

The UVOT instrument on the {\it Swift} satellite is executing a long-term program of 
observing many nearby SNe~Ia in the UV and optical 
(Brown et al. 2009, 2010, 2012; Milne et al. 2010, 2013; Foley et al. 2012a;  
Wang et al. 2009a; Bufano et al. 2009). 
With thousands of epochs of observations of more than 50 SNe~Ia, 
the UV emission from a large sample of  
events is being studied to search for universal characteristics as well as for patterns 
in the variations that are seen. Milne et al. (2013) found that the UV-optical colors 
of normal SNe~Ia 
evolve dramatically with epoch and that there exist two major color groups within the 
subset of normal SNe~Ia. The larger, NUV-red group contained roughly two-thirds of the 
sample, while a second NUV-blue group was offset from the NUV-red group, evolving bluer 
by 0.4 mag in $u-v$ (Figure \ref{u_v_color_curve}). The grouping was found to not correlate with 
peak-width. Although the differences were seen in all 
UV filters, in this work we will concentrate on the $u-v$, $u-b$ and $b-v$ colors. 
The ``irregular" and ``MUV-blue" minor groups discussed in M13 will be treated as 
NUV-red in this work. The ``irregular" group was concluded to be the optically 
broad-peaked subset of the NUV-red group, and the ``MUV-blue" group had colors similar 
to NUV-red SNe in the $u-v$ and $uvw1-v$ filter pairs.  

Since rest-frame $u$ and $v$ band emission can be observed in the optical for SNe~Ia at 
redshifts of $z\geq0.2$, we utilize four studies that have obtained spectra that span 
rest-wavelength range 3000--5700\AA.  E08 utilized the Low Resolution 
Imaging Spectrometer (LRIS) on the Keck I telescope to obtain 36 host-galaxy subtracted 
spectra of 36 SNe~Ia at 0.2$\leq$$z$$\leq$0.8. Twenty-three of those spectra span 
3000\AA--5700\AA\ and are included in this work. The fully-reduced and subtracted 
spectra were obtained from the Wiezmann Interactive Supernova data Repository, WISeREP 
(http://www.weizmann.ac.il/astrophysics/wiserep/). 
B09 utilized the FORS1 and FORS2 spectrometers to obtain 139 host-galaxy subtracted 
spectra of 124 SNLS SNe~Ia. Ten of these spectra span 3000\AA--5700\AA\ and are included in this 
work. The 10 spectra are in the redshift range, 0.38$\leq$$z$$\leq$0.52.  
The fully-reduced and subtracted ``snonly" spectra were obtained from the WISeREP 
site. F12 utilized the LRIS
spectrometer on the Keck I telescope to obtain single-epoch spectra of host-galaxy subtracted 
spectra of 21 SNe~Ia at 0.11$\leq$$z$$\leq$0.37. Nineteen of those spectra span 
3000\AA--5700\AA\ and the fully-reduced and subtracted spectra are included in this work.  
No spectra were rejected based upon reddening selection criteria, meaning that any reddening 
bias would have to have been employed by the sample selection of the E08, B09, F12 and R07 
publications. 
The UVOT-$v$ filter has some transmission redward of 5700\AA, but we make no correction for 
the range of maximum wavelengths; the maximum wavelength is simply required to be 5700\AA. 

\begin{figure}[h]
\epsscale{1.0} \plotone{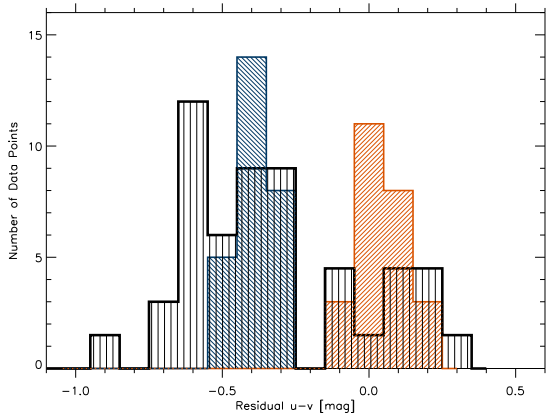}
\caption{Histogram of residuals of low- \& mid-$z$ $u-v$ photometry versus a linear fit to the UVOT 
NUV-red photometry. UVOT NUV-red photometry is shown in red with diagonal lines. UVOT NUV-blue 
photometry is shown in blue with blue diagonal lines. Mid-$z$ spectrophotometry is shown with 
thick, black vertical line, scaled up by 1.5. The low- \& mid-$z$ histograms peak at similar colors, with a 
switch from more NUV-red photometry at low-$z$ to more NUV-blue spectrophotometry at mid-$z$.
Linear fit was for $|$ t - t$_{BPEAK}$ $|$  $\leq$ 6 days.}
\label{u_v_histo}
\end{figure} 

The mid-$z$ spectra were folded through the UVOT $u$ and $v$ transmission curves to generate 
spectrophotometry (Tables 1 \& 2), 
utilizing the zero-points from Breeveld et al. (2012).\footnote{For the 
$ubv$ filters, the Breeveld et al. (2012) zeropoints are the same as Poole et al. (2008).} 
Figure~\ref{u_v_color_curve} 
shows the spectrophotometry compared to the UVOT photometry for the 
low-$z$ sample presented in M13. The mid-$z$ data overlaps the 
UVOT photometry well, and exhibits a separation into two groups as seen in the UVOT photometry. 
Interestingly, whereas two-thirds of the UVOT sample were NUV-red SNe~Ia, there are more NUV-blue 
SNe~Ia in the mid-$z$ sample. A linear fit to the color evolution of the UVOT NUV-red data for 
epochs $|t|$ $\leq$ 5 days allows color evolution with time to be removed, leaving only residual 
magnitudes relative to the fit. Figure \ref{u_v_histo} shows a 
histogram of the residuals, separating the UVOT data into NUV-red/blue groups, 
and leaving the mid-$z$ data as a single collection. The 
NUV-red/blue separation is clear from the histogram, and is reproduced by the mid-$z$ sample. The 
two peaks of the two mid-$z$ sample align with the two UVOT peaks, suggestive of similar colors 
within each group. If there is no systematic, erroneous 
color offset between the derivation of photometry from the 
low-$z$ and mid-$z$ samples, the alignment of peaks is suggestive of a lack of color evolution 
with redshift within the respective red and blue groups. The mean colors of the samples are shown 
in Table 3, showing that there is a consistent separation between the NUV-red and -blue groups, and 
that the mean colors within a group exhibit no change in color with redshift. 

\begin{table*}
\scriptsize
\vspace{-9mm}
\caption{Colors of mid-$z$ SN Ia Samples}
\begin{tabular}{l|c|cccc|cc|c|ccc}
\hline
\hline
SN & Sample$^{a}$ & Epoch & $z$ & s/$\Delta^{c}$ & E(B-V)/{\bf c}$^{d}$ & 
    $\lambda$(min) & $\lambda$(max) & v(SiII)$_{0}$ & $u-v$ & $u-b$ & $b-v$ \\
Name &          & [days]$^{b}$ &  [km/s] &  & [mag]  & [mag] &
 [Ang.]$^{e}$ & [Ang.]$^{f}$ & [mag]$^{g}$ & [mag]$^{g}$ & [mag]$^{g}$ \\
\hline
\multicolumn{11}{c}{NUV-Blue}  \\
\hline
03D1au & E08 & -1.6 & 0.504 & 1.13(02) & {\bf 0.02(03)} & 2476.2 & 5936.3 & 
      & -0.63 & -0.56 & -0.07  \\  
03D1dj & E08 & -1.8 & 0.400 & {\bf  ---} & {\bf  ---} & 2675.0 & 6389.3 & 
     & -0.63 & -0.46 & -0.16  \\  
03D3aw & E08 & -0.8 & 0.449 & 1.066(33) & {\bf -0.06(04)} & 2584.5 & 6162.9 & 
     & -0.83 & -0.48 & -0.35  \\  
03D3ay & E08 & -1.0 & 0.371 & 1.054(30) & {\bf -0.018(36)} & 2728.1 & 6521.3 & 
     & -0.54 & -0.48 & -0.06  \\  
%03D3bb & E08 &  2.2 & 0.244 & {\bf ---} & {\bf ---} &
%                        2995.1 & 7176.2 & -0.38 & -0.43 &  0.05  \\  
03D3bh & E08 & -3.5 & 0.249 & {\bf ---} & {\bf ---} & 2991.4 & 7152.0 & 
-11.5697  & -0.73 & -0.65 & -0.08  \\  
03D3cc & E08 &  9.1 & 0.463 & {\bf ---} & {\bf ---} & 2567.2 & 6122.2 &  
     & 0.13 & -0.18 &  0.31  \\  
03D3cd & E08 & -6.2 & 0.461 & 1.19(09) & {\bf -0.00(05)} & 2567.3 & 6130.6 & 
    & -0.84 & -0.74 & -0.10  \\  
03D4ag & E08 & -5.3 & 0.285 & 1.09(02) & {\bf -0.03(02)} & 2911.2 & 6974.4 & 
-9.44703  & -0.93 & -0.82 & -0.12  \\  
03D4cj & E08 & -7.4 & 0.270 & 1.12(01)  & {\bf -0.05(02)} & 2944.9 & 7047.2 & 
    & -0.83 & -0.69 & -0.14  \\  
04D1hd & E08 &  1.7 & 0.369 & 1.07(01) & {\bf -0.06(02)} & 2736.0 & 6527.6 & 
-11.1098 & -0.57 & -0.55 & -0.02  \\  
04D1pg & B09 & -1.3 & 0.515 & 1.10(02) & {\bf 0.12(04)} & 2811.0 & 5860.5 & 
      & -0.62 & -0.47 & -0.14  \\  
04D1rh & E08 &  2.7 & 0.435 & 1.11(03) & {\bf -0.02(03)} & 2599.5 & 6226.9 & 
    & -0.43 & -0.47 &  0.04  \\  
04D2fp & B09 &  1.8 & 0.415 & 1.03(02) & {\bf -0.00(03)} & 3010.5 & 6296.7 & 
    & -0.38 & -0.30 & -0.08  \\  
04D2gc & E08 & -0.2 & 0.522 & 1.12(03) & {\bf 0.05(04)} & 2457.9 & 5875.4 & 
      & -0.44 & -0.51 &  0.07  \\  
04D4in & E08 & -5.1 & 0.516 & 1.17(02) &  {\bf -0.04(03)} & 2463.7 & 5890.5 & 
     & -0.93 & -0.82 & -0.11  \\  
04D4jr & B09 & -5.9 & 0.470 & 1.15(02) & {\bf -0.02(03)} & 2896.9 & 6039.7 & 
    & -0.68 & -0.60 & -0.08  \\  
04D4jr & E08 & -0.1 & 0.482 & 1.15(02) & {\bf -0.02(03)} & 2520.2 & 6029.0 & 
     & -0.67 & -0.57 & -0.10  \\  
2005ik & F12 &  3.0 & 0.320 & 0.872(68) & 0.03(10) & 2521.9 & 7001.9 & 
    & -0.33 & -0.45 &  0.12  \\  
2005ix & F12 &  2.9 & 0.260 & 0.971(39) & 0.01(01) & 2740.5 & 7335.3 & 
-13.1845  & -0.44 & -0.39 & -0.05  \\  
2005jc & F12 &  0.8 & 0.215 & 0.995(27) & 0.12(04) & 2739.9 & 7607.0 & 
-10.6360  & -0.38 & -0.33 & -0.06  \\  
2005jd & F12 &  2.7 & 0.320 & 1.090(52) & 0.01(01) & 2659.9 & 7001.9 & 
-10.8679  & -0.36 & -0.37 &  0.01  \\  
2005ji & F12 &  1.0 & 0.214 & {\bf -0.07(09)} & 0.01(02) & 2775.1 & 7613.3 & 
-9.82174  & -0.62 & -0.49 & -0.12  \\  
2005jl & F12 & -0.1 & 0.180 & 1.003(32) & 0.10(03) & 2783.9 & 7832.7 & 
-10.2749  & -0.69 & -0.68 & -0.01  \\  
2005jn & F12 &  3.3 & 0.330 & 1.025(56) & 0.01(01) & 2557.1 & 6949.3 & 
-8.84449  & -0.73 & -0.74 &  0.01  \\  
2005jo & F12 & -1.0 & 0.230 & 1.055(39) & 0.14(07) & 2932.5 & 7514.3 & 
-9.54432  & -0.54 & -0.35 & -0.20  \\  
05D1ix & E08 & -9.1 & 0.490 & 1.06(02) & {\bf -0.04(03)} & 2503.4 & 6003.4 & 
   &  -0.70 & -0.48 & -0.22  \\  
05D1hk & E08 & -8.6 & 0.263 & 1.15(02) & {\bf 0.01(03)} & 2957.0 & 7065.9 & 
   & -0.75 & -0.78 &  0.03  \\  
05D1iy & E08 & -8.1 & 0.248 & {\bf ---} & {\bf ---} &  2997.3 & 7164.6 & 
   & -0.61 & -0.75 &  0.14  \\  
05D2ac & B09 &  2.0 & 0.479 & 1.11(02) & {\bf -0.01(03)} & 2879.5 & 6003.2 & 
    & -0.20 & -0.31 &  0.11  \\  
05D2bv & B09 & -0.1 & 0.474 & 0.99(01) & {\bf -0.09(03)} & 2889.6 & 6023.9 & 
     & -0.63 & -0.46 & -0.18  \\  
05D2dw & B09 & -5.3 & 0.417 & 1.11(02) &  {\bf 0.02(03)} & 3005.3 & 5736.4 & 
    & -0.78 & -0.53 & -0.25  \\  
05D4cw & B09 &  7.0 & 0.375 & 0.91(01) & {\bf -0.10(03)} & 3097.4 & 6457.4 & 
     & -0.15 & -0.15 & -0.00  \\  
2006pf & F12 &  2.0 & 0.366 & {\bf -0.32(10)} & 0.01(01) & 2358.6 & 6775.4 & 
    & -0.61 & -0.82 &  0.21  \\ 
2006pq & F12 & -4.4 & 0.193 & {\bf -0.36(13)} & 0.04(04) & 2651.2 & 7755.7 & 
-10.8753  & -0.77 & -0.66 & -0.11  \\  
2007lu & F12 & -0.1 & 0.319 & {\bf -0.47(07)} & 0.17(04) & 2434.4 & 6975.4 & 
-11.7068 & -0.41 & -0.51 &  0.10  \\  
2007lw & F12 &  1.5 & 0.290 & {\bf -0.23(09)} & 0.01(02) & 2569.7 & 7130.6 & 
-10.8118 & -0.54 & -0.46 & -0.09  \\  
2007qu & F12 & -1.2 & 0.310 & {\bf -0.25(14)} & 0.01(01) & 2466.4 & 7009.6 & 
-10.4310  & -0.77 & -0.79 &  0.02  \\  
\hline
\end{tabular}
\begin{tabular}{l}
$^{a}$ E08= Ellis et al. (2008); B09=Balland et al. (2009); F12=Foley et al. (2012).  \\
$^{b}$ Epoch of spectrum relative to B-band maximum. \\
$^{c}$ $\Delta$ or stretch (s) parameter for SN light curves. Stretch is from 
Guy et al. (2010), $\Delta$ (bold font) is from F12. \\
$^{d}$ E(B-V) or color (c) parameter for SN light curves. Color (c:bold font) is from 
Guy et al. (2010). \\
$^{e}$ Minimum rest-frame wavelength of spectrum. \\
$^{f}$ Maximum rest-frame wavelength of spectrum. \\
$^{g}$ Color of spectrophotometry. 
\label{tab1}
\end{tabular}
\end{table*}

\begin{table*}
\scriptsize
\caption{Colors of mid-$z$ SN Ia Samples (continued)}
\begin{tabular}{l|c|cccc|cc|c|ccc}
\hline
\hline
SN & Sample$^{a}$ & Epoch & $z$ & s/$\Delta^{c}$ & E(B-V)/{\bf c}$^{d}$ & 
    $\lambda$(min) & $\lambda$(max) & v(SiII)$_{0}$ & $u-v$ & $u-b$ & $b-v$ \\
Name &          & [days]$^{b}$ &     &  [km/s] &  [mag]  & [mag] &
 [Ang.]$^{e}$ & [Ang.]$^{f}$ & [mag]$^{g}$ & [mag]$^{g}$ & [mag]$^{g}$ \\
\hline
\multicolumn{11}{c}{NUV-Red}  \\
\hline
03D3af & E08 &  3.0 & 0.532 & {\bf ---} & {\bf ---} & 2444.5 & 5829.0 &  
     & 0.22 & -0.09 &  0.31  \\  
03D3bl & E08 &  3.8 & 0.355 & 1.02(02) & {\bf 0.24(03)} & 2759.5 & 6596.3 & 
-11.2125   &  0.20 &  0.00 &  0.20  \\  
04D2gc & B09 & -4.9 & 0.521 & 1.12(03)  & {\bf 0.05(04)} & 2800.1 & 5857.3 & 
    & -0.29 & -0.36 &  0.08  \\  
04D3ez & E08 &  1.5 & 0.263 & 0.89(01) & {\bf 0.07(03)} & 2961.2 & 7074.4 & 
-10.6360 & -0.07 & -0.22 &  0.15  \\  
04D3fk & E08 & -6.6 & 0.357 & 0.95(01)  & {\bf 0.10(02)} & 2754.5 & 6587.9 & 
-12.7075 & -0.30 & -0.34 &  0.05  \\  
04D4ju & B09 & -2.2 & 0.472 & 1.05(02) & {\bf 0.18(03)} &  2893.2 & 6031.7 & 
    & -0.21 & -0.25 &  0.04  \\  
2005jm & F12 & -4.6 & 0.204 & 1.079(39) & 0.17(09) & 2809.8 & 7676.5 & 
-12.0008 & -0.39 & -0.35 & -0.04  \\  
2005jp & F12 & -1.3 & 0.213 & 1.017(43) & 0.23(04) & 2779.5 & 7620.8 & 
-13.7803   & 0.02 & -0.15 &  0.17  \\  
05D2mp & E08 & -4.3 & 0.354 & 1.14(03) & {\bf 0.05(04)} & 2762.8 & 6596.7 & 
-12.2951 & -0.55 & -0.55 &  0.00  \\  
2006pt & F12 & -1.9 & 0.299 & {\bf -0.34(17)} & 0.08(11)  & 2436.4 & 7122.8 &  
-11.5844  & -0.06 & -0.22 &  0.28  \\  
06D4cq & B09 & -1.4 & 0.411 & 1.04(01) & {\bf -0.01(02)} & 3017.5 & 5760.3 & 
    & -0.29 & -0.20 & -0.09  \\  
2007ml & F12 & -4.0 & 0.190 & {\bf -0.08(09)} & 0.17(04) & 2706.7 & 7729.9 & 
-12.4129 & -0.20 & -0.21 &  0.01  \\  
\hline
\multicolumn{8}{c}{Undetermined}  \\
\hline
2006pz & F12 &  5.6 & 0.325 & {\bf -0.25(14)} & 0.08(08)  & 2560.7 & 6983.0 &  
    & 0.23 &  0.04 &  0.19  \\  
2005jk & F12 &  6.3 & 0.190 & 0.927(35) & 0.22(03) & 2846.2 & 7766.8 &  
-11.7950  & 0.35 &  0.16 &  0.18  \\  
03D3ba & E08 & 12.8 & 0.291 & {\bf 1.09(02)} & {\bf 0.14(04)} & 2896.5 & 6916.0 &  
     & 0.71 &  0.22 &  0.49  \\  
\hline
\end{tabular}
\begin{tabular}{l}
$^{a}$ E08= Ellis et al. (2008); B09=Balland et al. (2009); F12=Foley et al. (2010). \\
$^{b}$ Epoch of spectrum relative to B-band maximum. \\
$^{c}$ $\Delta$ or stretch (s) parameter for SN light curves. Stretch is in bold font. \\
$^{d}$ E(B-V) or color (c) parameter for SN light curves. Color (c) is in bold font. \\
$^{e}$ Minimum rest-frame wavelength of spectrum. \\
$^{f}$ Maximum rest-frame wavelength of spectrum.\\
$^{g}$ Color of spectrophotometry. \\
\end{tabular}
\label{tab2}
\end{table*}

There are a number of factors that could cause biases in these comparisons. Imperfect 
host-galaxy contamination removal could produce a spectrum attributed to the SN, but 
brighter or fainter in some part of the spectral range. All three spectral samples utilized 
photometry to estimate the SN multi-band brightness against which the final spectra could 
be compared. Further, F12 compared the spectra of the removed emission with observed galaxy 
spectra, arriving at solid matches. The SNLS samples, from E08 and B09, also utilized 
photometry of the SN and host galaxy via the PHASE technique to extract the SN component from 
the reduced spectrum (see B09 for details of the method \& Baumont et al. 2008). 
The level of agreement between the 
3 samples suggest that host galaxy contamination does not affect the determination of 
NUV-optical color group determination. 
As a check of the generation of UVOT spectrophotometry from spectra, we compare UVOT 
$u$ \& $b$ band photometry with spectrophotometry from HST spectra for 8 SNe~Ia that were 
observed both with UVOT and HST. The HST spectra were presented in Maguire et al. (2011). 
The statistical weighted mean difference in $u-b$ was 0.04~$\pm$~0.08 mag, due to 
$\Delta~u$=0.02~$\pm$~0.12 mag, and $\Delta~b$=-0.02~$\pm$~0.08 mag.  
We conclude that there is no significant offset between the HST spectrophotometry and the 
UVOT photometry. 
 
\begin{figure}[h]
\epsscale{1.0} \plotone{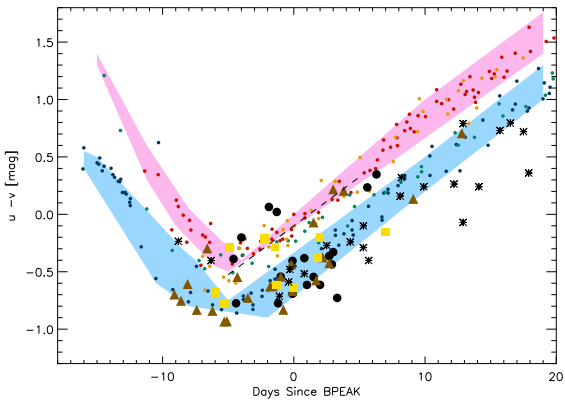}
\caption{$u-v$ colors of low-$z$, mid-$z$ \& high-$z$ samples. 
UVOT photometry is shown with small circles. 
Foley et al. (2012) high-$z$ spectrophotometry is shown with large, black circles. Ellis et al. (2008) 
high-$z$ spectrophotometry is shown with large, brown triangles. Balland et al. (2009) mid-$z$ 
spectrophotometry is shown with large, yellow squares. Riess et al. (2007) HST photometry is shown with 
black stars.  UVOT photometry is color-coded according to 
M13, with NUV-blue (blue), NUV-red (red), MUV-blue (orange) and irregular (green). The high-$z$ 
photometry exhibits similar color curves, as seen in the low- and mid-$z$ photometry.}
\label{r07_u_v_color_curve}
\end{figure}

SNLS-04D2gc appears in both groups, based upon two different spectra, with the pre-peak 
spectrum suggesting NUV-red and the at-peak spectrum suggesting NUV-blue. One NUV-blue event 
from M13, SN~2008hv, was red enough at similar epochs to appear in the NUV-red group, so perhaps 
SNLS-04D2gc and SN~2008hv feature a similar anomaly in the early-epoch evolution, and belong in 
the NUV-blue group. Alternatively, as the two spectra were obtained by different searches and 
were thus subjected to different host galaxy emission removal, it remains possible that the 
subtraction methodology introduced a bias. 
We note that the maximum wavelength for both spectra fail to reach the edge 
of the $v$-filter, leaving open the possibility of missed emission.  We include two SN~Ia 
spectra that were excluded from E08 analysis SNLS-05D1hk \& 03D4cj. 
Both are SN~1991T-like and appear to be similar to the other NUV-blue 
events at their respective epochs.\footnote{SNLS-03D3bb was excluded from this sample, 
as it is a super-Chandrasekhar candidate. 
It also appears to be similar to NUV-blue SNe~Ia, as was the super-Chandrasekhar candidate, SN~2009dc, 
featured in M13. Brown et al. (2014) concentrates on UVOT-observed super-Chandrasekhar candidates.} 
It is important to 
point out that the mid-$z$ histograms are derived from one data point per SN, while many data 
points per SN can be included in the UVOT low-$z$ histograms. 
The low-$z$ UVOT and HST NUV-red/blue ratios are taken from M13.  

\begin{center}
\begin{table*}
\scriptsize
\vspace{-9mm}
\caption{Mean Color Residuals of Samples}
\begin{tabular}{l|cc|cc|cc}
\hline
\hline
 & \multicolumn{2}{c}{$\Delta(u-v)^{a}$} &  \multicolumn{2}{c}{$\Delta(u-b)^{a}$} & 
                                                        \multicolumn{2}{c}{$\Delta(b-v)^{a}$} \\
 & \multicolumn{2}{c}{[mag]} &  \multicolumn{2}{c}{[mag]} & \multicolumn{2}{c}{[mag]} \\
\hline
           & low-$z$ & mid-$z$ & low-$z$ & mid-$z$ & low-$z$ & mid-$z$ \\
\hline
NUV-Red    & 0.04$\pm$0.09  & 0.09$\pm$0.14 & 0.19$\pm$0.07 & 0.17$\pm$0.09 & -0.00$\pm$ 0.04  & 
                                                                                 0.05$\pm$ 0.10  \\      
                                                                             
NUV-Blue  & -0.38$\pm$ 0.06 & -0.48$\pm$ 0.15 & -0.15$\pm$ 0.07 & -0.21$\pm$ 0.15 & -0.10$\pm$ 0.08 & 
                                                                                  -0.12$\pm$ 0.11 \\
\hline
\end{tabular}
\begin{tabular}{l}
$^{a}$ Residual colors relative to fits described in the text. \\
\end{tabular}
\label{tab3}
\end{table*}
\end{center}

Riess et al. (2007) obtained HST photometry and spectra of high-$z$ SNe~Ia with redshifts exceeding 
$z~\geq~0.6$. None of the spectra in that study span the $u-v$ wavelength range, but the 
K-corrected $U$ and $V$ photometry can be compared to the UVOT photometry. 
Figure \ref{r07_u_v_color_curve}
shows that the high-$z$ HST photometry matches the NUV-blue group with nearly all 
eligible SNe~Ia NUV-blue 
(2002eb, HST05Gab, 2003xx, HST040mb, HST04Rak, 2002fw, 2003eq, HST05Dic, HST05Str). 
As few as zero or as many as two SNe from this sample could be NUV-red; 
%Only 0-2 SNe~Ia are NUV-red, 
we list SN~2002hr as NUV-red and SN~2002kd as undetermined.   
This suggests an even stronger dominance of NUV-blue events at high redshifts. The K-corrected 
photometry was used as published, with no attempt to generate separate NUV-blue and NUV-red 
K-corrections. 

\begin{figure}[h]
\epsscale{1.0} \plotone{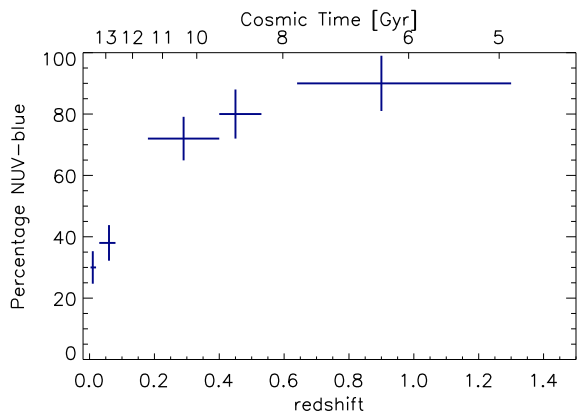}
\caption{Percentage of NUV-blue SNe~Ia versus all SNe~Ia as a function of redshift. At low-$z$, there are more 
NUV-red events, but with increasing redshift, the NUV-blue events become the dominant group.}
\label{blue_red_ratio}
\end{figure}

Dividing the mid-$z$ sample into two bins based on redshift, we see that there is a transition 
from NUV-red dominance at low-$z$ to NUV-blue dominance at high-$z$ (Figure \ref{blue_red_ratio}). 
The UVOT and HST low-$z$ SN groupings are from M13. 
The mid-$z$ groups are binned as 0.18 $\leq$ $z$ $\leq$ 0.4 and 0.4 $\leq$ $z$ $\leq$ 0.53, 
respectively, with 29 and 20 SNe~Ia in the two mid-$z$ bins. There are 
23 UVOT SNe~Ia, 23 HST low-$z$ SNe~Ia and 10 HST high-$z$ SNe~Ia, as shown in Table 4. 
A more quantitive discussion of UV dispersion is presented at the end of Section 3. 

The shift from dominance of NUV-red events at low-$z$ to dominance of NUV-blue events at mid-$z$ is 
suggestive of a fundamental change in the properties of SNe~Ia with redshift. Both the 
%Ellis et al. (2008)  
Maguire et al. (2012) and the F12 low/mid-$z$ spectral studies reported that the 
UV-optical colors were bluer for the mid-$z$ sample. The photometry comparisons suggest that the 
reason for the bluer colors at mid-$z$ is the dominance of NUV-blue SNe~Ia at mid-$z$ instead of 
NUV-red SNe~Ia. It is also quite reasonable to understand that generating a mean spectrum from 
a combination of NUV-red and -blue SNe~Ia will lead to the UV wavelength range featuring much 
more scatter. Two samples which have mean colors that differ by 0.4 mag are being treated as a 
single group in those studies.

\begin{table}[t]
\scriptsize
\vspace{-9mm}
\caption{Change with redshift of NUV-red versus NUV-blue ratio}
\begin{tabular}{l|cc|cc}
\hline
\hline
SN Sample & N(Red)$^{a}$ & N(Blue)$^{a}$ & \% Blue & $\Delta$(\% Blue)$^{b}$ \\
\hline
 & & & & \\
UVOT               & 16 & 7  & 30 & 11 \\
low-$z$ HST        & 15 & 8  & 35 & 10 \\
mid-$z$ $\leq$ 0.4 & 8  & 21 & 72 & 14 \\
mid-$z$ $\geq$ 0.4 & 4  & 16 & 80 & 20 \\
high-$z$           & 1  & 9  & 90 & 27 \\
\hline
\end{tabular}
\begin{tabular}{l}
$^{a}$ Number of SNe~Ia in sample determined to be \\
NUV-red or NUV-blue.\\
$^{b}$ Poisson error of percentages, additionally accounting \\
for SNe with uncertain determinations.\\
\end{tabular}
\label{tab4}
\end{table}

One alternative explanation to NUV-blue SNe~Ia being the dominant explosion variety at 
higher redshifts is that the NUV-blue events are more luminous and are over-represented in 
a magnitude limited sample. There are a number of factors that argue against that explanation. 
First, none of the samples were rest-frame $u$-band dominated when determining candidates, 
so a luminosity bias would only occur if NUV-blue SNe were optically bright. The fact that 
this separation was not recognized years ago argues against a difference of more than 
0.1 - 0.2~mag. Further, E08 and F12 studied magnitude selection effects and determined that their 
samples were not strongly biased. Second, all three mid-$z$ surveys showed the same NUV-blue 
dominance, despite the differences between the sample selection for each survey. 
Third, even if there were a luminosity bias of ~0.3~mag, the 
effect would be much too small to shift the NUV-blue:red ratio from 40\% to 90\%. 
We conclude that there is a real variation in the NUV-blue to NUV-red ratio with
redshift.

\section{Spectral Comparisons of NUV-red and NUV-blue SNe~Ia}

\begin{figure*}[h]
%\vspace{-8mm}
\epsscale{1.80} \plotone{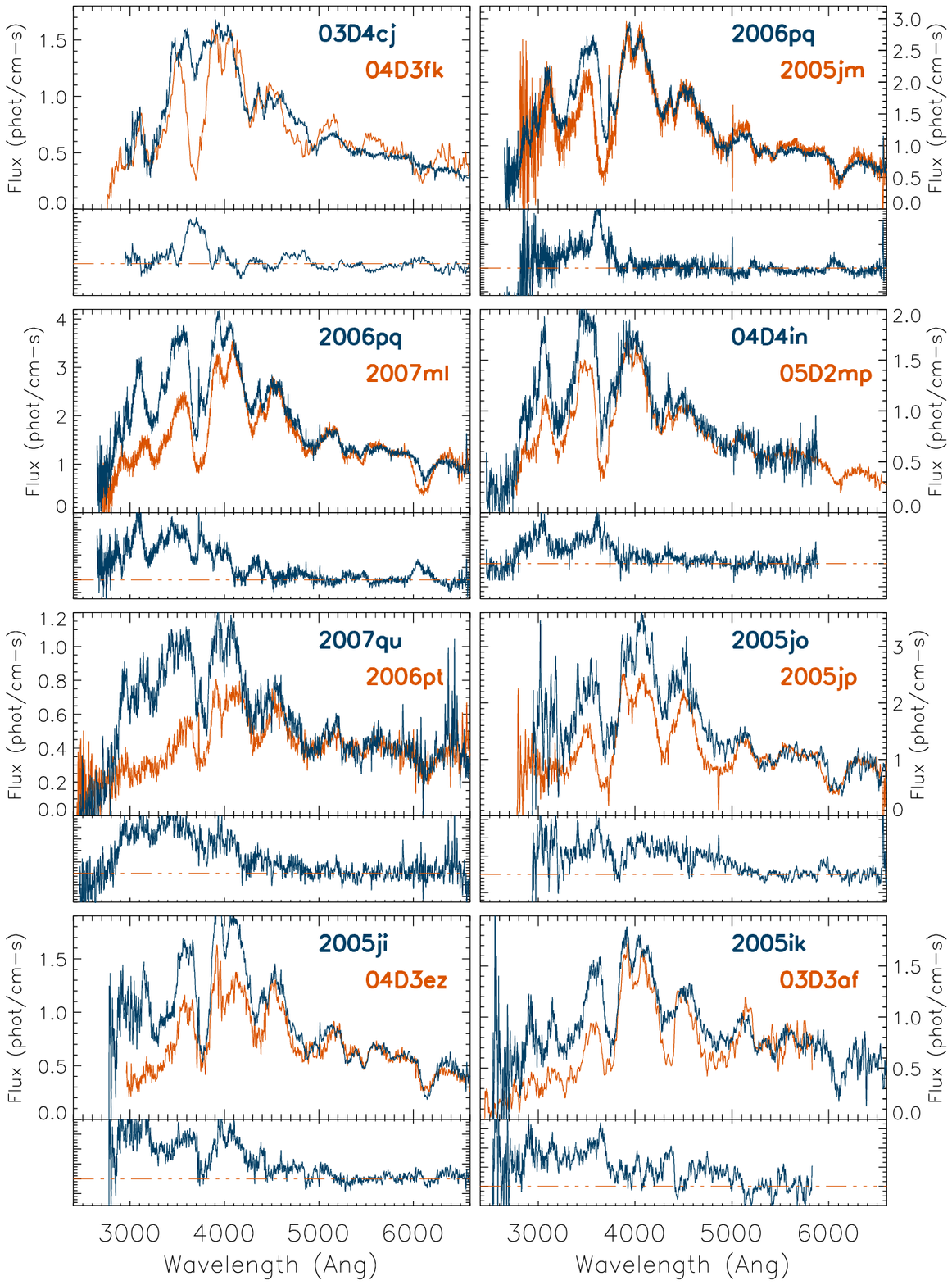}
\caption{Comparisons of NUV-blue and -red SN spectra from Ellis et al. (2008), 
Balland et al. (2009) and Foley et al. (2012) mid-$z$ samples. Solid blue lines are 
NUV-blue events, red-dashed lines are NUV-red events. The spectra are normalized to the 
overlap in the 5000 -- 6000$\AA$ wavelength range, or to the red edge of the spectrum.
Below each spectrum is the residual spectrum of the NUV-blue minus NUV-red. Spectra 
were matched in epoch and peak width. Normalizations that do not reach 6000$\AA$ are 
apparent in the residual spectra.}
\label{compspec_mixed}
\end{figure*}

Using the same approach as employed for the low-$z$ \textit{HST} SN
spectra in M13, we compare spectra of NUV-red/blue pairs.  For two SNe
to be considered ``pairs,'' we require that they 
%come from the same
%data sets (to minimize any differences in sample bias), 
have similar
optical light-curve shapes (in the same third of the sample), 
have similar phases (within 1 day excepting one pair), 
and have {\it different} UV colors.  
%From
%the SN pairs, we then select pairs of spectra, requiring that the
%spectra be of SN pairs and have similar phases.  
Selecting the best matches that meet these requirements
allow a direct comparison of SN spectra which largely removes possible
differences attributable to light-curve shape (and thus luminosity)
and phase; therefore, any differences in the spectra are more likely
to be related to the differences in UV colors.  We present these 
spectra pairs in Figure~\ref{compspec_mixed}.

As expected, the spectra of SNe from the NUV-blue group have a higher
UV flux than those of SNe from the NUV-red group.  Interestingly,
there is no noticeable difference in the optical continua ($\lambda
\gtrsim 4500$~\AA) of SNe from the two groups.  This is perhaps not
unexpected given the similar optical

\noindent
colors for all SNe.  But it
indicates that this effect is not caused by odd reddening, which would
have some subtle difference in the optical continuum, or the
confluence of a strong continuum difference offset by differences in
spectral features.

Also notable is that the same UV spectral features are present for
both groups.  Although the strengths of these lines are different for
each SN, it is clear that the UV color difference is the result of a
difference in a wide band in the UV wavelength range and not the 
presence/absence of particular lines or dramatically different 
line strengths (and unaffected UV continua) in one particular group.

The final clear difference in the two groups is the velocity of
multiple spectral features.  When the \ion{Si}{2} $\lambda 6355$
feature is observed with a reasonable signal-to-noise ratio, the
NUV-red SN often has a broader, higher-velocity line than the NUV-blue
paired SN (and the NUV-blue spectrum never has a higher-velocity
feature for our pairs). 
This is sometimes easier to see in the residual spectra, which
shows a positive bump at $\sim$ 6100~\AA, indicating excess absorption
in the NUV-red spectra.  This is perhaps best seen in the comparison
of SNe~2006pq and 2007ml.  Since the SNe have similar light-curve
shapes and the spectra are at similar phases, we conclude that the
velocity of the \ion{Si}{2} $\lambda 6355$ feature is linked to the UV
colors.  We note that the trend with velocity is not exclusive to
\ion{Si}{2}; in particular, \ion{Ca}{2} H\&K shows similar behavior.

%\citet{Foley11:vel} 
Foley \& Kasen (2011) first showed that the intrinsic $B_{\rm max} -
V_{\rm max}$ color of a SN~Ia was strongly linked to the velocity of
the \ion{Si}{2} $\lambda 6355$ feature near maximum light.  The
physical explanation for this effect is that higher-velocity spectral
features are also broader, 
creating additional 
%and thus a SN photosphere had additional
opacity in the blue where line blanketing dominates, while further to
the red, electron scattering dominates the opacity, and thus the
broader lines do not increase the overall opacity for those
wavelengths.  This interpretation predicts that other than the obvious
velocity differences, spectral differences between low and
high-velocity SNe should be largely constrained to the bluer
wavelengths, that the differences should largely be a difference in
the continuum, and that the difference should persist to the UV.
Figure~\ref{compspec_mixed} supports this interpretation, and the
differences in UV colors may predominantly be caused by different
opacities resulting from different ejecta velocities. With a larger 
UVOT sample, an interesting 
comparison will be between Low Velocity (LV) NUV-blue SNe and LV NUV-red 
SNe. If the \ion{Si}{2} velocity distributions are similar between those 
groups, the prediction would be that the $b-v$ colors would be similar. 

Thomas et al. (2011) reported a correlation between UVOT NUV-blue
SNe~Ia and the detection of an absorption feature at $\lambda$
6580~\AA\ attributed to \ion{C}{2}.  M13 confirmed the correlation for
all seven UVOT NUV-blue SNe in the sample.  Since the $\lambda$ 6580\AA\ 
feature is only detected at early epochs ($t \lesssim -4$~days) in
normal SNe~Ia, one would only expect to detect this feature in
early-epoch spectra that reach to 6650~\AA.  Six mid-$z$ spectra meet
those criteria.  The comparison spectra of the NUV-blue SN~2006pq
shows a notch at the location of the \ion{C}{2} feature, when compared
with the NUV-red SN~2007ml, but in general the spectra either cut-off
before the \ion{C}{2} feature, were of too low S/N, or were obtained
at too late an epoch for \ion{C}{2} identification.  There are
therefore too few SNe from which we can draw strong conclusions.
However, several groups have found that high-velocity SNe~Ia, which we
associate with NUV-red objects, are far less likely to have detectable
\ion{C}{2} (Parrent et al. 2011; Folatelli et al. 2011; 
Silverman \& Filippenko 2012), consistent with the findings of M13.

F12 generated mean spectra of low-$z$ and mid-$z$
samples and determined that when normalized to the optical emission,
the mid-$z$ mean spectrum featured a UV excess, the same result that was found 
by Maguire et al. (2012).  Recognizing the
increased fraction of NUV-blue events at mid-$z$, and the similarities
of the distributions of low-$z$ and mid-$z$ colors within the red/blue
groups (shown in Figure~\ref{u_v_histo}), the comparison of mean
spectra can be revisited with the additional separation between
NUV-red and -blue SNe.  We note that several of the spectra used here
were also used by F12.

\begin{figure}[h]
\epsscale{1.0} \plotone{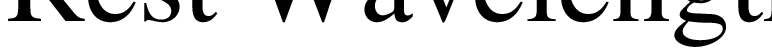}
\caption{Mean spectra from high-$z$ NUV-red SNe~Ia compared with the mean low-$z$ spectrum. 
The red spectrum is the mid-$z$ NUV-red mean spectrum, the black spectrum is the mean low-$z$ 
spectrum (F12). The spectral samples are limited to near-peak epochs, 
$|$ t$_{BPEAK}$ $|$ $\leq$ 7 days, and are normalized over the 4500\AA $\leq$ 7500 \AA wavelength 
region.The two spectra approximately agree.}
\label{highz_red_vs_lowz}
\end{figure}

\begin{figure}[h]
\epsscale{1.0} \plotone{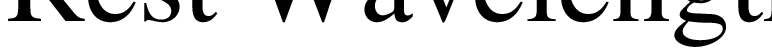}
\caption{Mean spectra from mid-$z$ NUV-blue SNe~Ia compared with the mean low-$z$ spectrum. 
The blue spectrum is the mid-$z$ NUV-blue mean spectrum, the black spectrum is the mean low-$z$ 
spectrum (F12). The spectral samples are limited to near-peak epochs, 
$|$ t$_{BPEAK}$ $|$ $\leq$ 7 days, and are normalized over the 4500\AA $\leq$ 7500 \AA 
wavelength region.
The NUV-blue spectrum exhibits a UV excess relative the the mean low-$z$ spectrum.}
\label{highz_blue_vs_lowz}
\end{figure}

\begin{figure}[h]
\epsscale{1.0} \plotone{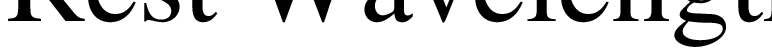}
\caption{Mean spectra from high-$z$ SNe~Ia separated into NUV-red and NUV-blue groups. 
The red spectrum is the high-$z$ NUV-red mean spectrum, the blue spectrum is the mid-$z$ 
NUV-blue mean spectrum. 
The spectral samples are limited to near-peak epochs, $|$ t$_{BPEAK}$ $|$ $\leq$ 7 days, 
and are normalized over the 4500\AA\ $\leq$ 7500 \AA\ wavelength region. The 
NUV-blue mean spectrum is bluer in the UV wavelength.}
\label{highz_red_blue}
\end{figure}

We used the methods of Foley et al. (2008) and F12 
to generate mean spectra and uncertainty spectra corresponding to the
boot-strap resampling uncertainty.  Using only spectra near peak
brightness ($|$ t$_{BPEAK}$ $|$ $\leq$ 7~days) and normalizing in the
rest-frame wavelength region $4500 \leq \lambda \leq 7500$~\AA, mean
spectra for the low-$z$ sample \citep{foley12c} and
mid-$z$ NUV-red and NUV-blue samples are presented in
Figures~\ref{highz_red_vs_lowz} to \ref{highz_red_blue}.

Comparing the mean mid-$z$ NUV-red and low-$z$ mean spectra
(Figure~\ref{highz_red_vs_lowz}), we see that the spectra are nearly
identical.  There are two clear differences: The mid-$z$ NUV-red has
broader and higher-velocity \ion{Si}{2} $\lambda$6355 and Ca H\&K
features.  The difference in the Ca H\&K feature is exclusively to the
blue portion of the complex feature \citep{Foley_Kirshner_2013}, which has a significant
contribution from \ion{Si}{2} $\lambda$3727 (but also see
Childress et al 2013).  As the low-$z$ sample has both
NUV-red and NUV-blue SNe, one can extrapolate from the previous
comparison of pairs that the relatively small difference in velocity
is the result of the NUV-blue SNe in the low-$z$ sample.

Another difference in the mid-$z$ NUV-red and low-$z$ mean spectra is
the strength of the 4800~\AA\ feature, with the mid-$z$ NUV-red mean
spectrum having a weaker absorption than that of the low-$z$ mean
spectrum.  \citet{foley08} reported a similar difference in a
comparison of low-$z$ and mid-$z$ mean spectra.  This feature is
predominantly caused by \ion{Fe}{2} absorption with additional
absorption from \ion{Si}{2} and \ion{Fe}{3}.  Because of the blending
of this feature, there are multiple plausible explanations for such a
difference ranging from excitation/temperature differences to
differences in composition.

Examining Figure~\ref{highz_blue_vs_lowz}, we see more dramatic
differences between the mid-$z$ NUV-blue and low-$z$ mean spectra.
The mean mid-$z$ NUV-blue spectrum has lower velocity features, a UV
excess, and weaker spectral features compared to the mean low-$z$
spectrum.  Specifically, the mid-$z$ NUV-blue mean spectrum has excess
flux compared to the low-$z$ mean spectrum for wavelengths shorter
than $\sim$ 4000~\AA.  This is consistent with what has been seen in
other comparisons (Cooke et al. 2011, Maguire et al. 2012, F12) and
supports the suggestion that the spectral evolution with redshift is
dominated by the change in membership from NUV-red events at low-$z$
to NUV-blue events at high-$z$. 

Most spectral features in both the UV and optical, including the Ca
H\&K feature, the ``\ion{Mg}{2} 4300'' feature, and the 4800~\AA\ 
feature, are weaker in the mid-$z$ NUV-blue mean spectrum.  In
retrospect, a similar behavior can be seen between the
intermediate-redshift and low-redshift mean spectra of F12. 
Considering that many of the same spectra
contribute to the F12 mean spectra and the mean
spectra presented here, this observation is reassuring.  Intriguingly,
the lower 
%($\mean{z} = 0.19$) and higher ($\mean{z} = 0.31$) 
($\overline{z} = 0.19$) and higher ($\overline{z} = 0.31$)
redshift
subsamples presented by F12 also show this trend with
the higher redshift mean spectrum having muted features relative to
the lower redshift mean spectrum.

We compare the mean mid-$z$ NUV-blue and NUV-red spectra in
Figure~\ref{highz_red_blue}; as expected, there are significant
spectral differences: the NUV-blue spectrum has lower velocity
features, a UV excess, and weaker spectral features.  Since the
mid-$z$ NUV-red mean spectrum was similar to the low-$z$ mean
spectrum, most of the above discussion between the mid-$z$ NUV-blue
mean spectrum and the low-$z$ mean spectrum also applies to this
comparison.

The wavelength of maximum absorption for the \ion{Si}{2} $\lambda$6355
feature corresponds to a blueshifted velocity, $v_{Si~II}$, of
$-10$,$960 \pm 160$~\kms and $-12$,$280 \pm 80$~\kms, for the NUV-blue and
NUV-red mean spectra, respectively.  For a subset of the individual
mid-$z$ spectra, we were able to measure $v_{Si~II}$, and its value at
$B$-band maximum brightness, $v_{Si~II}^{0}$ \citep{Foley11:vgrad}.
The NUV-blue and NUV-red samples had average $v_{Si~II}^{0}$ of
$-10$,630 and $-11$,850~\kms, respectively.  Similarly, the two groups
had median values of $-10$,620 and $-11$,740~\kms, respectively.  The
velocity difference in the mean spectra are duplicated in measurements
of individual objects.  All measurements suggest that the NUV-red
sample has an average ejecta velocity that is $\sim$12\% 
larger than that of the NUV-blue sample.\footnote{The \ion{Si}{2} $\lambda$6355 
line velocities are all blueshifted, and thus negative, but ``above" and 
``below" will be relative to the absolute magnitude of the velocities.}

Of the 15 NUV-blue SNe for which we could determine $v_{Si~II}^{0}$,
14 have velocities below $-11,800$~\kms, the nominal separation
between ``high-velocity'' and ``low-velocity'' SNe~Ia
\citep{Foley11:vel}.  In contrast, 4 of the 8 NUV-red SNe for which
we could measure $v_{Si~II}^{0}$ had a velocity higher than
$-11,800$~\kms.  M13 found similar results for the low-$z$ sample: the
NUV-blue objects exclusively had low-velocity ejecta, while half of
the NUV-red objects were found in each group.  Performing a
Kolmogorov--Smirnov test, we find that the NUV-blue and NUV-red SNe
likely have different parent populations with respect to ejecta
velocity ($p = 0.0027$). By contrast,  performing a
Kolmogorov--Smirnov test on the stretch values, we find that the 
NUV-blue and NUV-red SNe likely have similar distributions of stretch 
values ($p = 0.87$).

We also examined the pseudo-equivalent width (pEW) of the \ion{Si}{2} $\lambda 4131$ feature, 
which has been proposed as an indicator of \ion{Si}{2} velocity and color, and is a possible way 
to further reduce Hubble scatter 
\citep[e.g.,][]{Arsenijevic08, Blondin_etal_2011, Nordin_etal_2011}.  Specifically, 
SNe with higher \ion{Si}{2} $\lambda 4130$ pEW tend to have higher maximum-light velocities and 
redder colors, as one would expect given the velocity-color relation \citep{Foley11:vel}.  
For the low-$z$, NUV-Blue, and NUV-Red composite spectra, we measure pEWs of 8.0, 6.4, and 11.4~\AA, 
respectively, with an estimated systematic uncertainty of $\sim$ 2.0~\AA\ that greatly dominates over 
any statistical uncertainty.  This measurement is consistent with NUV-Blue SNe being intrinsically 
bluer and having lower eject velocities than NUV-Red SNe.

Reconcentrating on the general features of the mean spectra, 
the mid-$z$ NUV-blue and NUV-red mean spectra clearly differ in the NUV when 
normalized in the optical wavelength range. The differences invite the question 
as to whether the higher dispersion reported in mean UV spectra is wholly due 
to failing to separate the SNe~Ia into red and blue groups. Comparing the 
dispersion in an optical 
band (4500--5500\AA) with a NUV band (3000--3500\AA) for the low-$z$ mean spectrum, 
we find the NUV dispersion to be 4.2 times the optical dispersion. By constrast, that 
ratio is 2.8 for the mid-$z$ NUV-blue mean spectrum and 2.7 for the  mid-$z$ NUV-red 
mean spectrum. Stated a different way, the NUV-blue and NUV-red mean spectra retain 
68\% and 63\% of the NUV dispersion after separation compared to the low-$z$ (unseparated) 
mean spectrum. This suggests that there remains intrinsic dispersion within each group that 
is independent of NUV-red/blue group membership, but that recognizing the group membership 
makes the NUV wavelength range closer to being useful relative to the optical wavelength range. 

\section{Comparing $u-b$ and $b-v$ colors of low- and high-$z$ Samples}

\begin{figure}[h]
\epsscale{1.0} \plotone{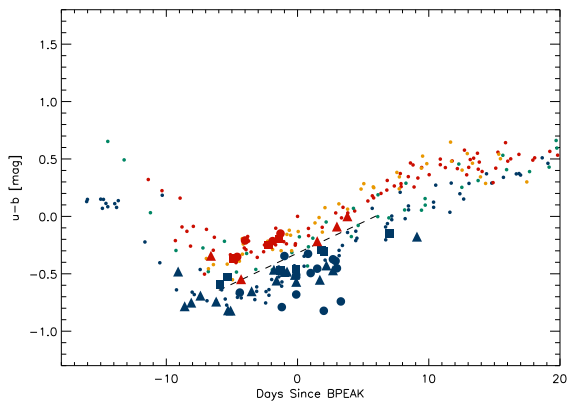}
\caption{$u-b$ colors of low-$z$ versus mid-$z$ samples. Figure symbols are similar to 
Fig. 3 except the mid-$z$ SNe~Ia are color coded red or blue to match 
NUV-red/blue determinations from $u-v$ spectrophotometry.}
\label{u_b_color_curve}
\end{figure}

\begin{figure}
\epsscale{1.0} \plotone{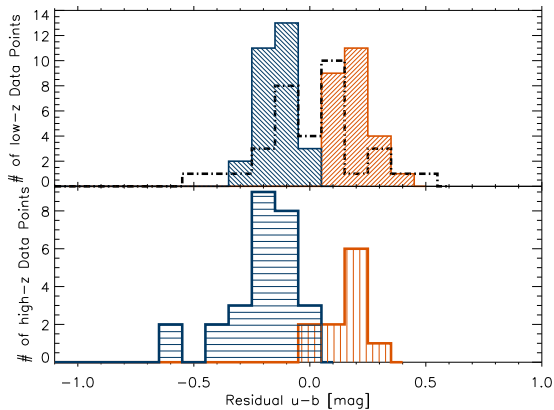}
\caption{Histogram of residuals of low- \& mid-$z$ $u-b$ photometry versus a linear fit to the 
UVOT NUV-red photometry offset to the blue by 0.2 mag. The upper panel shows UVOT 
NUV-red photometry (red), UVOT NUV-blue photometry (blue) and HST spectrophotometry from M13 
(black-dashed). The lower panel shows 
mid-$z$ spectrophotometry with NUV-red SNe~Ia (red/vertical) and NUV-blue SNe~Ia (blue/horizontal). 
The NUV-red/blue groups are separated at both redshifts, but with less separation than the 
$u-v$ photometry. The low- \& mid-$z$ histograms peak at similar colors. 
Linear fit was for $|$ t - t$_{BPEAK}$ $|$ $\leq$ 6 days.}
\label{u_b_histo}
\end{figure}

Utilizing the $u-v$ colors to determine NUV-red versus NUV-blue membership allows the 
differences in the $u-b$ and $b-v$ colors to be studied. Figure \ref{u_b_color_curve} 
shows the $u-b$ color curves of the mid-$z$ sample compared to UVOT photometry. The figure 
is similar to Figure \ref{u_v_color_curve}, but the mid-$z$ spectrophotometry has been 
color-coded red or blue based upon the NUV-red/blue determinations shown in Table~1.  
A linear fit to the NUV-red UVOT photometry is shown as a dashed line, as was performed in 
M13. The line was fitted for epochs $|$ t $|$ $\leq$ 6 and offset to the blue by 0.2 magnitudes 
so that there is a sign change between NUV-red/blue groups.\footnote{The linear fit was not 
started at earlier epochs to avoid the abrupt early color change. The linear fit was not 
extended to later epochs to avoid NUV-blue/irregular confusion, as discussed in M13.} 
Histograms of the residuals to that fit are shown 
in Figure \ref{u_b_histo}. There is still a color difference between the two groups, but it 
is less than for the $u-v$ colors. Table 2 shows the lower separation between NUV-red/blue, 
which lowers from 0.42 and 0.57 mag to 0.34 and 0.38 mag , respectively for the low-$z$ and mid-$z$ 
samples, 
but again shows no evolution of color with redshift within a group. 

\begin{figure}[t]
\epsscale{1.0} \plotone{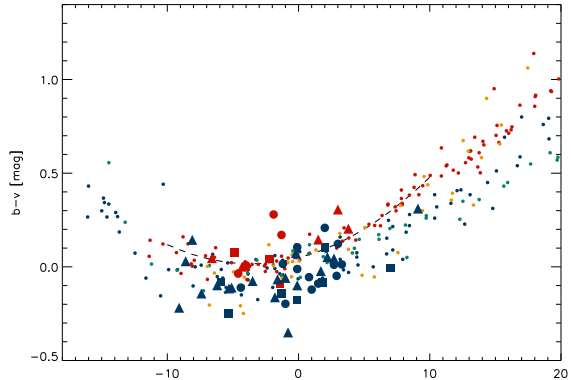}
\caption{$b-v$ colors of low-$z$ versus mid-$z$ samples. Figure symbols are similar to 
Fig. 3}
\label{b_v_color_curve}
\end{figure}

\begin{figure}
\epsscale{1.0} \plotone{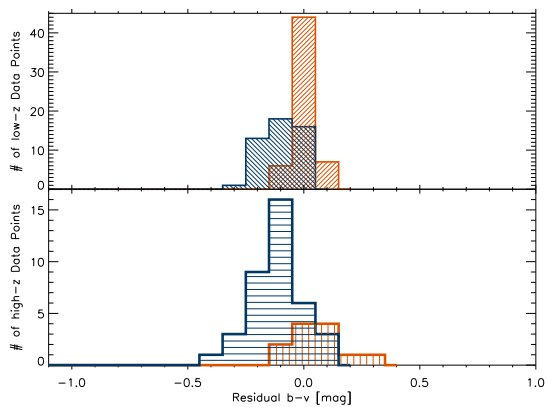}
\caption{Histogram of residuals of low- \& mid-$z$ $b-v$ photometry versus a quadratic fit to the 
UVOT NUV-red photometry. The upper panel shows UVOT 
NUV-red photometry (red) and UVOT NUV-blue photometry (blue). The lower panel shows 
mid-$z$ spectrophotometry with NUV-red SNe~Ia (red/vertical) and NUV-blue SNe~Ia (blue/horizontal).  
The NUV-red/blue groups are not separated at either redshift, but the NUV-blue samples are bluer with 
overlap with the NUV-red samples. The low- \& mid-$z$ histograms peak at similar colors. 
Quadratic fit was for $|$ t - t$_{BPEAK}$ $|$ $\leq$ 10 days.}
\label{b_v_histo}
\end{figure}

The same method can be employed to study $b-v$ colors. Figures \ref{b_v_color_curve} and  
\ref{b_v_histo} show the color curves and resulting histograms for the $b-v$ colors. A 
quadratic fit is utilized for the $b-v$ photometry, using data during the 
$|$ t $|$ $\leq$ 10 day epoch. 
The NUV-red and NUV-blue distributions are less separated, with overlap for both 
samples. Table 2 again shows the lower separation between NUV-red/blue, 
which lowers to 0.10 and 0.17 magnitudes for the low-$z$ and mid-$z$ samples,
but yet again shows no evolution of color with redshift within a group. 

\section{Cosmological Implications of NUV-red/blue $b-v$  Color Differences}

The color difference between the $b-v$ colors of the NUV-red and -blue
groups at both redshifts are potentially problematic for the cosmological
utilization of SNe~Ia as distance indicators. The optical colors,
particularly the $B-V$ color, is used to estimate the host-galaxy
extinction to the SN. Current implementations of $B-V$ colors do not
differentiate NUV-red from NUV-blue SNe~Ia. As these methods are
trained wholly or primarily on nearby, low-$z$ SNe, they are likely to be dominated by
NUV-red SNe~Ia. An algorithm so trained would interpret the bluer
$B-V$ color of NUV-blue events to imply very low, even negative
extinction. If there really were host-galaxy extinction along the line
of sight to that SN, the under-estimation of that extinction would
lead to the erroneous impression that the SN is less luminous than
it really is.  
The change in the NUV-red/blue ratio in favor of NUV-blue events, combined 
with the under-estimation of extinction for NUV-blue events 
means that many high-$z$ SNe~Ia could be
systematically more luminous than recognized. It is thus critical to
determine if there are differences between the absolute magnitudes of
NUV-blue SNe~Ia and NUV-red SNe~Ia. With the recognition of the
existence of two groups, it becomes critical to accumulate a sample of
nearby NUV-blue and NUV-red SNe~Ia for which the absolute magnitudes can
be determined.  The lack of color changes with redshift within each
group suggests that a new generation of extinction estimation based on
low-$z$ SNe~Ia would be valid at high-$z$.

Without absolute magnitudes determinations that separate NUV-blue from 
NUV-red SNe~Ia, we do not know the exact bias that results from treating 
both groups as a single sample. However, making the assumption that the 
absolute optical magnitudes are the same, we can assess the impact of
this potential bias. To do this, we perform two similar simulations.  For both, we
assume that the two groups have a difference in $b-v$, $\Delta (b-v)$, 
that directly impacts the distance modulus such that $\Delta \mu =
\Delta A_{V} = R_{V} \Delta (b-v)$ (note that the sign of this value
is set this way because we defined $\Delta (b-v) > 0$).  The average
Hubble residual bias is then

\begin{align}
  {\rm HR} &= 0.5 * \Delta \mu/ (f_{\rm b} - f_{\rm r}) \\
%      <HR> &= 0.5 * \Delta \mu/ (f_{\rm b} - f_{\rm r}) \\
           &= 0.5 * \Delta \mu/ (2 f_{\rm b} - 1),
\end{align}

where $f_{\rm b}$ and $f_{\rm r}$ are the fraction of NUV-blue and
NUV-red SNe, and assuming that $f_{\rm b} + f_{\rm r} = 1$.  We then
use the fractions of NUV-blue and NUV-red SNe as determined in
Section~2 to determine the evolution of the Hubble residual
bias as a function of redshift.

The first simulation assumes $\Delta (b-v) = 0.10 \pm 0.09$ or $\Delta
(b-v) = 0.17 \pm 0.15$, the differences seen for the low-$z$ and
mid-$z$ samples, respectively.  We also assume that $R_{V} = 2$ or
$3.1$, resulting in four distinct possibilities for the Hubble
residual bias.  Since there is no specific preference for one
possibility over the others, we perform a Monte Carlo simulation
drawing from each of the four possibilities to determine the median
bias and the uncertainty on the bias.

The second simulation is similar to the first, but uses $\Delta (b-v)
= 0.059 \pm 0.013$.  This is the measured $B_{\rm max} - V_{\rm max}$
offset between the low and high-velocity subsamples of the
Foley \& Kasen (2011) sample.  As shown above, ejecta velocity is highly
correlated with NUV color.  Although this color difference may not be
exactly the difference between the NUV color subclasses, it will be
illustrative because of the proportionally small uncertainty for the
color difference.  For this simulation, we also use both values of
$R_{V}$ and perform a Monte Carlo simulation to determine the final
bias and uncertainty.

\begin{figure*}
\epsscale{2.0} \plotone{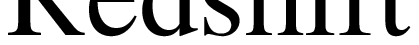}
\caption{Simulated Hubble residuals due to the variations with redshift of the NUV-red/blue ratio 
and the HV/LV ratio. The solid blue points show Hubble Residuals based upon the observed $b-v$ 
color difference between NUV-red and NUV-blue SNe~Ia. The solid black points show Hubble Residuals 
based upon the observed $b-v$ color difference between HV and LV SNe~Ia, offset by 0.01 in redshift for 
display purposes. See text for details of the simulations.}
\label{hr}
\end{figure*}

Figure~\ref{hr} shows the results of the simulations.  The Hubble
residuals shown are relative to a sample that has equal numbers of
NUV-blue and NUV-red SNe.  Both predict the Hubble residual bias to
change with redshift with the bias being negative at low $z$ and
positive at high $z$.  This is expected since NUV-red SNe, which
should have their extinction overestimated and their distance moduli
underestimated, are more prevalent in the local universe.

The magnitude of the bias depends on the magnitude of the $b-v$
difference.  The first simulation, which uses the measured difference
directly from the NUV subclasses indicates a bias of $HR = 0.15 \pm
0.11$ ($0.18 \pm 0.14$) between $z = 0$ and $z \approx 0.5$ ($z
\approx 1$), while the second simulation finds $HR = 0.07 \pm 0.02$
($0.09 \pm 0.03$) for the same redshifts.  Of course the second
simulation has a more statistically significant bias because of the
relatively smaller uncertainty for the $b-v$ difference.  Of particular
note is that given the current uncertainties, the first simulation is
consistent with zero bias for all redshifts.

\begin{figure*}
\epsscale{1.8}
%\plotone{smeargrid_MLCS2k2_fitres.ps}
%\includegraphics[width=\textwidth]{smeargrid_MLCS2k2_fitres}
\plotone{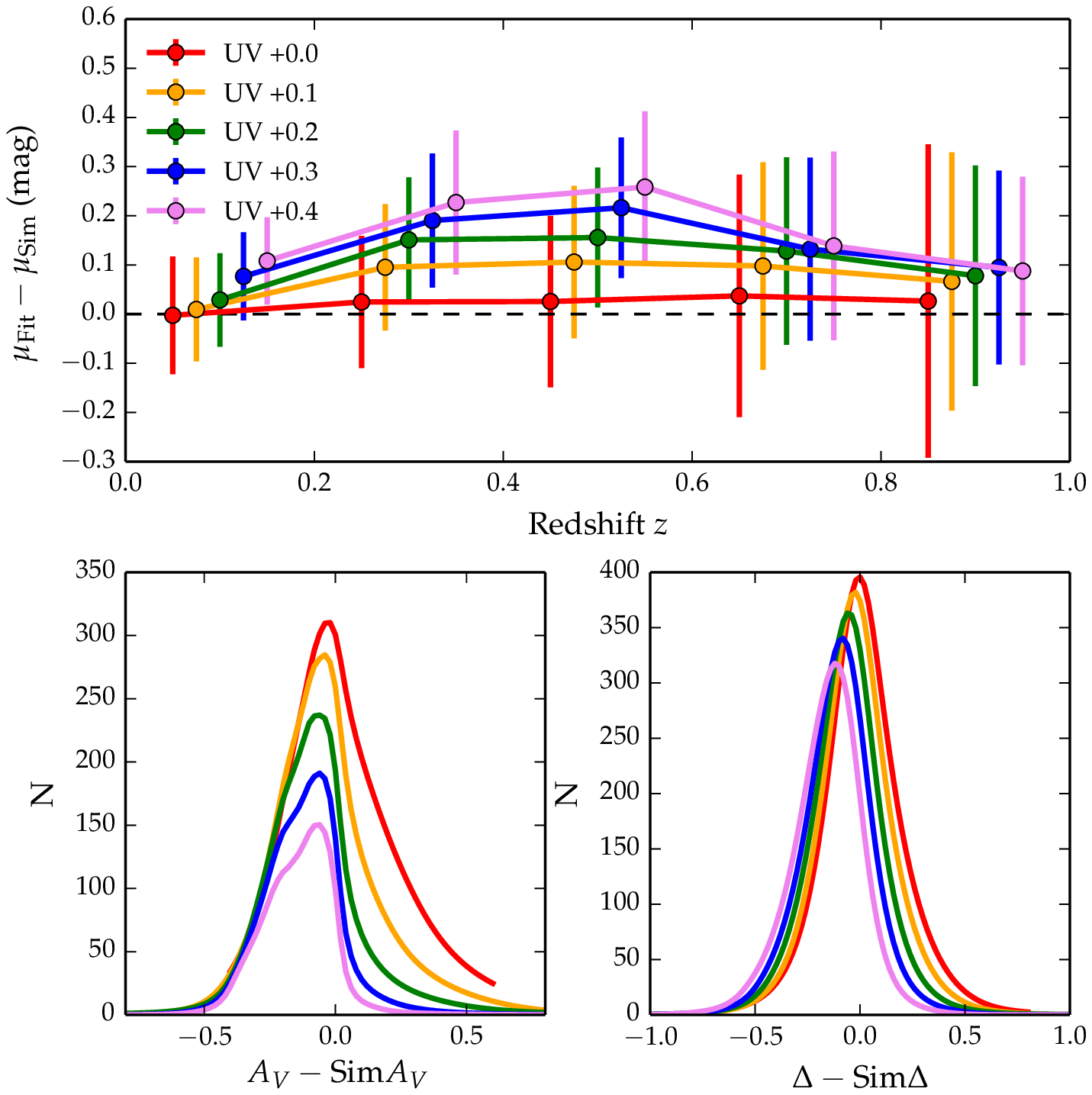}
\caption{Differences between fit and simulated light curve parameters with bias
increasing from $\Delta U = 0.0$~mag (red) to $\Delta U = -0.40$~mag (violet).
Increasing the NUV bias causes {\it MLCS2k2} to favor lower values of the
extinction $A_{V}$ and light curve shape parameter, $\Delta$ (lower panels) in
order to fit the brighter and bluer SNe~Ia light curves. As the distance
modulus, $\mu$ is negatively correlated with $A_{V}$, the light curve fitter
exhibits a systematic bias to higher distances. There is significant evolution
of the mean distance modulus bias with redshift (upper panel, shown using the
mean residual in redshift bins of 0.2, with different values of the bias offset
slightly along the x-axis for clarity) arising from a combination of effects.
The standard deviation of the residuals in each bin (rather than the error of
the mean) is shown in each bin, to indicate the range spanned by the data. Our
simulated light curves are `flux smeared' to replicate the observed scatter in
the Hubble diagram, with reasonable intrinsic dispersion.
} \label{fit_param_resid}
\end{figure*}

To investigate the potential effect of two UV-optical color groups on the 
$w$ parameter, 
we have modified the SNANA (v10.34) \citep{Kessler_etal_2009}
{\it SNLC-SIM} routines to allow us to incorporate a filter-dependent bias
in the rest-frame {\it MLCS2k2} model \citep{Jha_etal_2006} of the SN. As the Swift
$u$ band is considerably more blue than Johnson $U$, we have elected to
study the effect for a range of different values from $0.0$~mag (no bias or
NUV-Red) to $-0.4$~mag (strong blue NUV bias) in increments of $-0.1$~mag. To
reduce the number of independent parameters, we related the bias in Johnson $B$
to Johnson $U$ via $\Delta B = 0.4\Delta U$. As our analysis of Hubble
residuals previously indicated a strong trend with redshift, we have elected to
simulate all SNe~Ia in the simulation as originating from a single survey
modelled after the SNLS, to avoid conflating redshift-dependent trends arising
from different survey properties and selection effects with the signal from the
UV bias.

\begin{figure*}
\epsscale{1.8}
%\plotone{w-1.0_bias_effect.ps}
\plotone{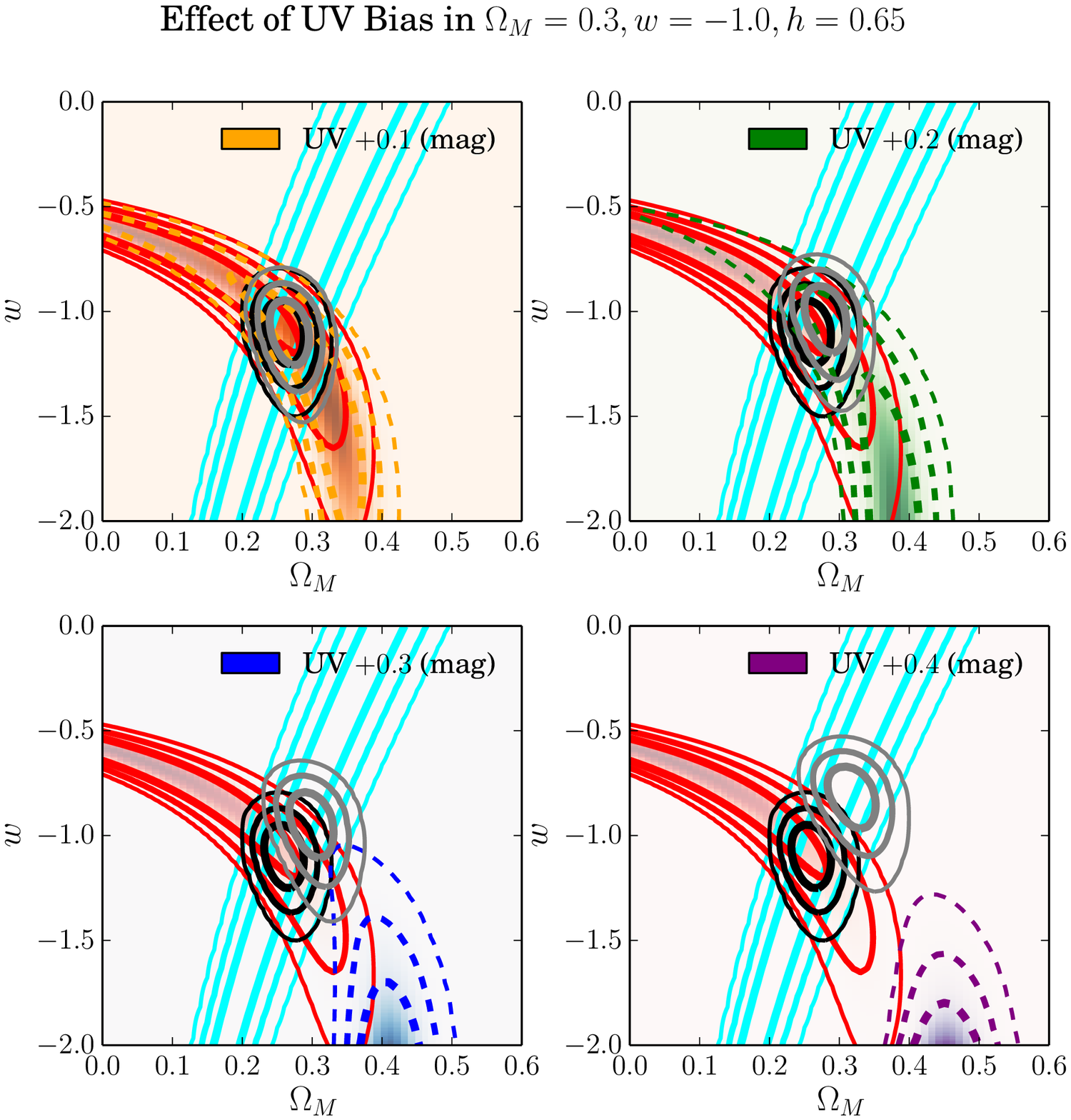}
\caption{Effect of increasing NUV bias on cosmological inference in a flat
$\Omega_{M}=0.3$, $w=-1$, $h_{0}=0.65$ cosmology. The BAO contours are shown in
cyan, while the unbiased, or NUV-Red sample is shown in red. The dashed
contours show the effect of combining the NUV-Blue and NUV-Red SNe~I for
different values of the bias (colors correspond to those shown in the legend
and Figure \ref{fit_param_resid}). As increased bias leads to a systematically
higher distance determinations, we find the SNe~Ia contours are systematically
biased to more negative values of the dark energy equation of state, $w$, and
higher $Omega_{M}$. However, the combined BAO and SNe~Ia constraints (shown in
back for the NUV-Red sample, and grey for the combined sample) are biased
towards lower values of $w$, and there is a strong increase in the tension
between the two datasets. These results indicate that values of the bias above
$\Delta U = -0.3$~mag are implausible. The tension is qualitatively similar to
the tension seen between the Union 2.1 supernova sample and the Planck
constraints.} \label{cosmo_fit} 
\end{figure*}

We simulate 500 SNe~Ia for each value of the bias, in a flat cosmology with
$\Omega_{M} = 0.3$, $h_{0} = 0.65$ and three different values of the dark
energy equation of state, $w = -1, -0.5, 0$. Additional flux smearing is used
to ensure the simulated light curves reproduce the observed scatter in the
Hubble diagram, and the cosmological fits have reduced $\chi^{2}$$\sim$$1$.
All SNe~Ia are fit with {\it MLCS2k2} with priors appropriate for the SNLS.
The effect of the NUV bias on the {\it MLCS2k2} fit parameters $\mu$, $A_{V}$,
and $\Delta$ are presented in Figure \ref{fit_param_resid}. {\it MLCS2k2}
interprets the excess blue color as reduced extinction, and the recovered
$A_{V}$ distribution becomes increasingly asymmetric and skewed to lower values
with increase in the bias.  However, as $A_{V}$ is negatively correlated with
the distance modulus, $\mu$, increasing NUV bias leads to an increasing positive
bias in the recovered distances. We find two effects with increasing redshift.
As rest-frame $I$ and $R$ redshift out of the observer frame $griz$ filters,
the relative weight of $U$ and $B$ increases. Additionally, at very high-$z$,
our simulation samples a very narrow range of low $A_{V}$ as we are unlikely to
find heavily extincted SNe~Ia. However as {\it MLCS2k2} does not allow
negative $A_{V}$ this leads to a reduced bias in the recovered extinction at
high-$z$. Consequently, the bias in recovered distance modulus first increases
up to  $z$$\approx$$0.5$ and then decreases. The recovered light curve shape
parameter, $\Delta$ shows an increasing bias to more negative, or broader,
intrinsically brighter models with increase in the strength of the NUV bias.
Our simulation uses a simple magnitude offset as the bias, and does not alter
light curve shape in any band, and we do not find any significant bias in the
recovered time of maximum, as we would expect.

\begin{figure*}
\epsscale{1.6}
%\plotone{w-1.0_1dmarg_bias_effect.ps}
%\includegraphics[width=\textwidth]{w-1p0_1dmarg_bias_effect}
\plotone{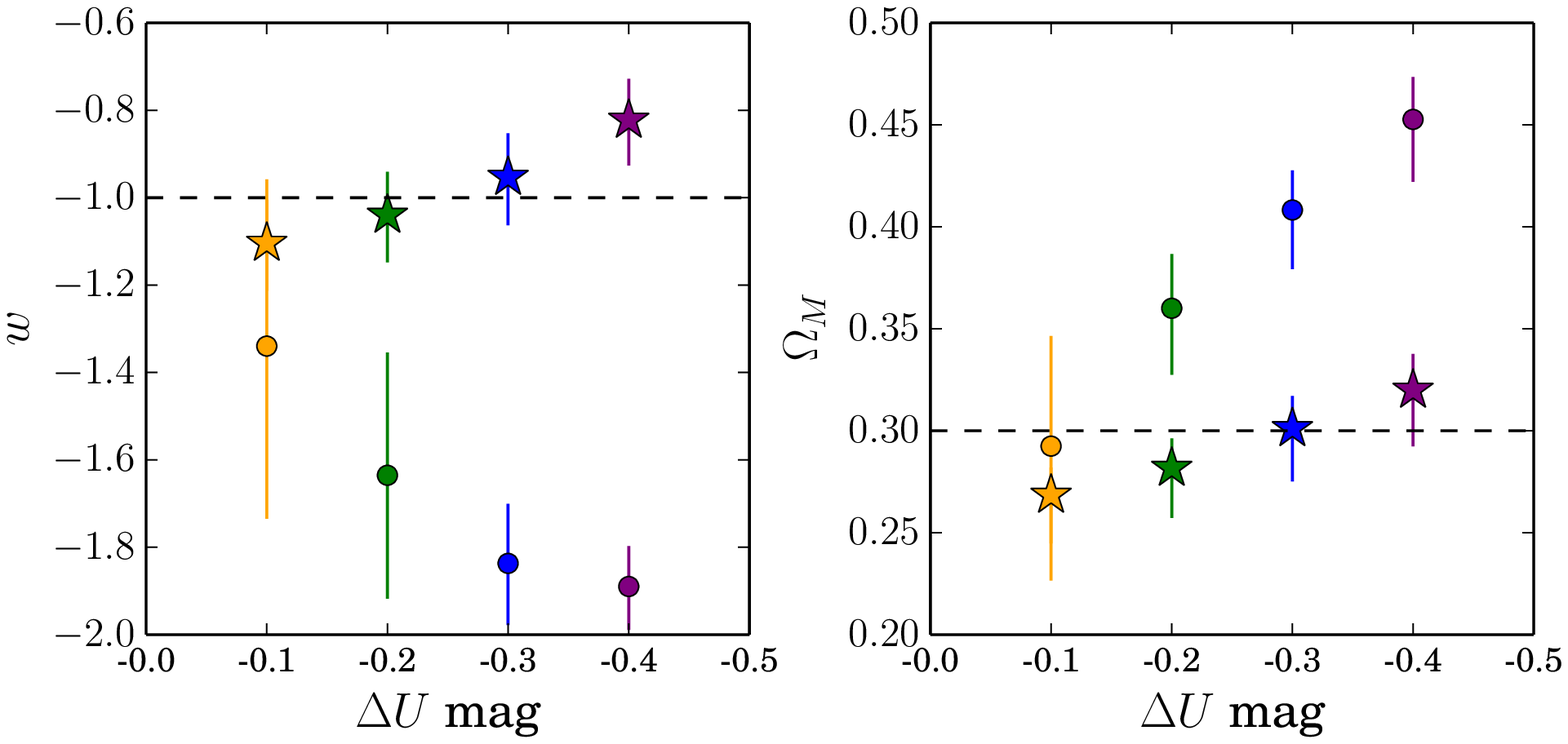}
\caption{1D marginalized values of $\Omega_{M}$ and $w$ as a function of the
bias for SNe~Ia only (circles) and joint SNe~Ia and BAO datasets (stars). There
are clear trends illustrating that the cosmological parameter bias increasing
from the SNe~Ia only as the NUV bias is increased. While the joint constraints
appear superficially to show much less bias, this is illusory as the tension
between the two datasets increases with bias two, as evidenced by the
divergence of the two trends. The equation of state of the dark energy, $w$ is
significantly more strongly impacted by increasing NUV bias. This is not
unexpected as the BAO data alone are a strong constraint on $\Omega_{M}$. }
\label{cosmo_resid}
\end{figure*}

For each value of the bias, we create a combined sample of unbiased, NUV-Red
SNe~Ia and NUV-Blue SNe~Ia consistent with Figure \ref{blue_red_ratio}. We use
the
{\it simple-cosfitter}\footnote{\url{http://qold.astro.utoronto.ca/conley/simple_cosfitter}}
(v1.6.11) package to estimate the cosmological parameters, $\Omega_{M}$ and $w$
for each combined sample, and compare the results to the cosmological
parameters estimated from the full unbiased NUV-red sample. The effect of the
NUV bias on the cosmological parameters is shown in Figure \ref{cosmo_fit}. As
increasing the NUV bias leads to an increased positive bias in recovered
distance modulus, $\mu$, we find a bias towards more negative values of the
equation of state parameter, $w$ from the SNe~Ia data. When combined with the
BAO prior, the joint constraints are biased towards lower values of $w$ and
higher values of $O_{M}$. However, the tension between our simulated SNe~Ia
data and the real BAO data also increases with increase in the bias. The
$99.7\%$ SNe~Ia contours do not overlap with the the BAO contours in the case
of the strong $\Delta U = -0.4$~mag bias, and the combined constraints are not
meaningful. It is likely that biases over $\Delta U = 0.3$~mag are unphysical.
The increased tension is not captured by the results from simple parameter
marginalization shown in Figure \ref{cosmo_resid}.

There are several limitations in these simulations  - given the small sample
size presented in this work, there is still considerable uncertainty about the
size of the difference between the NUV colors, the physical mechanism that
causes it, and the relative ratio of the two groups and it's evolution with
redshift. Additionally, we have made several simplifying assumptions such as
simulating a single survey, and relating the bias in Johnson $B$ to Johnson
$U$. More extensive analysis would be premature given the current sample size.
The quantitative results shown in Figure \ref{cosmo_resid} can only be
interpreted within the context of these limitations, and the qualitative
behavior seen with in light curve fit parameter residuals and the cosmological
contours arguably offers more insight.  In particular, the tension between our
simulated SNe~Ia data, and the real BAO constraints is similar to the tension
observed between constraints from Planck and the Union 2.1 supernova 
set \citet{Rest_etal_2013}. There
are many sources of systematics or potential new physics that could cause the
observed tension between the SNe~Ia data and the results from CMB experiments,
and we cannot definitely ascribe it to the effect of two groups of UV-optical
colors in SNe~Ia. As the size of the {\it Swift} SNe~Ia sample increases, and high-$z$
searches such as Pan-STARRS and DES probe the rest-frame UV with better S/N, it
will become possible to identify and characterize different color groups more
thoroughly.

%Nonetheless, a potential bias of 0.03 -- 0.3~mag would have significant 
%implications for cosmological results. 
%Further investigations are necessary to constrain the color
%difference, determine if the color difference evolves with redshift,
%and precisely determine the fractions of the NUV subclasses with
%redshift.

Absolute magnitudes of the early UVOT SNe~Ia were presented in Brown
et al. (2010). That sample was dominated by NUV-red events with a
single NUV-blue SN, 2008Q and used existing color relations and extinction laws.
Rather than extend that methodology to the current, larger sample, it
is important to wait until an improved treatment of intrinsic colors and 
extinction is developed, and present the absolute magnitudes of the UVOT
sample at that time. Efforts are also being made to observe
SNe Ia in the nearby Hubble flow with UVOT to reduce the effect of 
peculiar and/or thermal velocities on the redshift-derived distances.  For the 
more nearby SNe, improved distances using Cepheids or other distance indicators 
would allow the differences seen in the UV photometry as well as 
spectroscopy \citep{Foley_Kirshner_2013} to be compared directly to the 
absolute magnitudes in the optical.

\section{Summary}

The existence of two UV-optical color groups amongst normal SNe~Ia has
been established by UVOT photometric observations. Three mid-$z$
spectroscopic samples have had spectrophotometry performed to produce
UVOT $u$, $b$ and $v$ spectrophotometry. The three samples have been
combined and compared with the UVOT low-$z$ sample, finding the same
two groups at higher redshift. The colors within each group match the
low-$z$ colors, but the bluer, NUV-blue group dominates at high-$z$ (90\%),
the opposite of what is seen at low-$z$ ($\sim$33\%). 
This change appears to be the
explanation for the difference between high-$z$ mean spectra and
low-$z$ mean spectra that were generated from these samples. A higher
redshift sample exhibits an even stronger dominance of NUV-blue
events.

Separating the mid-$z$ sample into NUV-red/blue groups, we investigate
the $u-b$ and $b-v$ colors. The $u-b$ colors also exhibit a
NUV-red/blue separation, with the NUV-blue group again the bluer
group, but the two samples begin to overlap. The $b-v$ color feature
significant overlap, but the NUV-blue sample is bluer on average by 
0.1--0.2 mag. This
presents problems for methods of extinction estimation for SNe~Ia that
utilize $B-V$ colors, as existing methods do not differentiate
NUV-red/blue events. The use of SNe~Ia as cosmological distance indicators rely
upon distant SNe~Ia being similar to nearby SNe~Ia, at least to the
extent that there are nearby reference SNe~Ia for all distant SNe~Ia.
The finding that NUV-blue events dominate distant SNe~Ia means that
the absolute magnitudes, and a separate luminosity-width relation
needs to be derived for this group of normal SNe~Ia.

We found that the NUV color differences were highly correlated with
the measured ejecta velocity.  Optical color differences have also
been reported between normal SNe~Ia that have been divided into two
groups based upon the blueshift of a prominent \ion{Si}{2}
$\lambda$6355 absorption feature (\citealp{Foley11:vel,Foley11:vgrad}, F12).  
Foley \& Kasen (2011) suggested that the physical reason
for the correlation between color and ejecta velocity was driven by
the connection between the depth of the line-forming region within the
ejecta and the opacity.  One can interpret this connection as higher
ejecta velocities causing overlapping lines and thus increased opacity
and redder $B-V$ colors or as increased Fe-group abundance increasing
the opacity and causing the line-forming region for intermediate-mass
elements to be further out in the ejecta and thus at higher velocity.

M13 showed that NUV-blue SNe~Ia have low velocities while the NUV-red
SNe~Ia are equally mixed between low and high-velocity objects.
Approximately two-thirds of low-$z$ SNe~Ia have low-velocity ejecta
\citep[e.g.,][]{Foley11:vgrad}.  Meanwhile, about one-third of low-$z$
SNe~Ia are NUV-blue.  Low-$z$ SNe~Ia can be separated into two roughly
equal-sized groups: low-velocity/NUV-blue, low-velocity/NUV-red, and
high-velocity/NUV-red.  The low-velocity/NUV-blue events are bluer in
$B-V$, leading to the observed color differences.

This contrasts with the mid-$z$ sample, where about three-quarters of
the sample are NUV-blue.  For the subsample where we could determine
their ejecta velocities, the mid-$z$ NUV-blue SNe were essentially all
low-velocity, while half of the mid-$z$ NUV-red SNe are high-velocity.
Therefore, the three groups listed above, low-velocity/NUV-blue,
low-velocity/NUV-red, and high-velocity/NUV-red, have approximate
fractions of 76\%, 12\%, and 12\%, respectively.

We note that these fractions are seen for a subsample of mid-$z$
SNe~Ia, specifically those with relatively high-quality spectra.  As
detailed by E08 and F12 it is unlikely
that these subsamples had selection effects that cause this color
difference.  However, it is possible that the true fractions for the
entire SN~Ia population at higher redshifts could vary from those
reported here.

These findings motivate revisiting the SN templates that are used in 
comparisons with high-redshift SNe~Ia. Two sets of templates should be 
produced, one for NUV-blue SNe~Ia and one for NUV-red SNe~Ia. A key for  
inclusion into the new template sets will be whether NUV-red/blue 
membership can be determined. For archival SN~Ia observations, 
that would require that at least 
some observations of a given SN were performed with a detector with 
acceptable blue sensitivity, to allow clear separation between NUV-red and 
NUV-blue. For future SN~Ia observations, these findings suggest placing a 
high priority on detectors with excellent blue sensitivity, or building 
a template set from UVOT-observed SNe~Ia. An alternative possibility would be 
for the observations of known NUV-red and NUV-blue SNe to reveal another 
observable characteristic that completely correlates with the NUV-optical 
Along those same lines, it will be important to understand the associations 
between NUV-red/blue groupings and other characteristics that are independent 
of peak width, such as: HV/LV grouping, the presence or absence of unburned carbon 
in early-epoch spectra, whether the late $B$ and $V$ band photometry follows the 
Lira relation and whether the optical spectra show evidence of strong Na D 
absorption lines. Collectively, correlations between these groupings will be 
a powerful probe of the progenitor metallicity, the binary pair, and the 
explosion physics of the different varieties of SNe~Ia.

P.A.M. acknowledges support from NASA ADAP grant NNX10AD58G. 
P.J.B. is supported by the Mitchell Postdoctoral Fellowship and NSF grant AST-0708873.
All
supernova observers thank the mission operations team at Penn State for
scheduling the thousands of individual UVOT target-of-opportunity observations
that comprise the UVOT dataset. The NASA/IPAC Extragalactic Database (NED) was
utilized in this work. NED is operated by the Jet Propulsion Laboratory of the
California Institute of Technology, under contract with the National
Aeronautics and Space Administration.

\end{document}